\documentclass[twocolumn,preprintnumbers,epsfig,amsmath,amssymb]{revtex4}

\usepackage{graphicx}
\usepackage{dcolumn}
\usepackage{bm}

\begin{document}

\title{ Exact Analytical Solution of the Constrained Statistical Multifragmentation Model
and Phase Transitions in Finite Systems }  

\author{Kyrill A. Bugaev}

\affiliation{Bogolyubov Institute for Theoretical Physics,
Kiev, Ukraine\\
%
%
Lawrence Berkeley National Laboratory, Berkeley, CA 94720, USA
}

\date{\today}

\begin{abstract}
We discuss  an  exact  analytical
solution of 
a simplified version of the  statistical multifragmentation model with the  
restriction that the largest fragment size cannot exceed the finite volume of the system. 
A complete analysis of the isobaric partition singularities of this model is done for finite volumes. 
It is shown that the real part of  any simple pole of the isobaric partition defines the free energy
of the corresponding state, whereas its  imaginary part, depending on the sign,  defines
the inverse decay/formation time of this state. 
The developed  formalism allows us, for the first time, to exactly define the finite volume analogs
of gaseous, liquid and mixed  phases of this model  from the first principles of statistical mechanics and 
demonstrate the pitfalls of  earlier  works. 
The finite
size effects for large fragments and the role of metastable (unstable) states are discussed. 

\vspace*{0.2cm}

\noindent
\hspace*{0.0cm}{\bf PACS} numbers: 25.70. Pq, 21.65.+f, 24.10. Pa
\end{abstract}

\maketitle



\section{Introduction}

\vspace*{-0.25cm}

A great deal of progress was recently achieved in our understanding of the multifragmentation phenomenon 
\cite{Bondorf:95, Gross:97, Moretto:97} when an exact analytical solution of a simplified version of 
the statistical multifragmentation model (SMM) \cite{Gupta:98,Gupta:99} 
was found in Refs. \cite{Bugaev:00,Bugaev:01,Reuter:01}.  
An invention of a new powerful mathematical method \cite{Bugaev:04a}, 
the Laplace-Fourier transform, 
allowed us  not only  to solve this  version of SMM analytically for finite volumes \cite{Bugaev:04a}, but 
to find the surface partition and surface entropy  of large clusters for  a variety of statistical ensembles  \cite{Bugaev:04b}.
It was shown \cite{Bugaev:04a}
that for finite volumes the analysis of the grand canonical partition (GCP) of the simplified SMM
is reduced to the analysis of the simple poles of the corresponding isobaric partition,  obtained 
as a Laplace-Fourier transform of the GCP.
This method opens  a principally new  possibility 
to study the nuclear liquid-gas phase transition directly from the partition of finite system  and 
without  taking its thermodynamic limit.


Exactly solvable models with phase transitions play a special role in the statistical
physics - they are the benchmarks of our understanding of critical phenomena that occur  in 
more complicated  substances.
They are our theoretical laboratories, where we can study the most fundamental problems of critical phenomena
which cannot be studied 
elsewhere. 
Note that these questions {\it in principle} cannot be clarified either  within  the widely used 
mean-filed approach or numerically.   

Despite this success, the application of the exact  solution \cite{Bugaev:00,Bugaev:01,Reuter:01}
to the description of experimental data is  limited because this solution  
corresponds to an infinite system volume.
Therefore, from a practical point of view it is necessary to extend the formalism for finite volumes.
Such an extension is also necessary because, despite a general success in the understanding
the nuclear multifragmentation,  there is a lack of a systematic and rigorous  theoretical approach
to study the phase transition phenomena in finite systems. 
For instance, even the best   formulation of  the statistical mechanics and 
thermodynamics of finite systems  by Hill \cite{Hill} is not rigorous while discussing
the phase transitions.
Exactly solvable models of phase transitions applied to finite systems 
may provide us with the first principle results
unspoiled by the additional simplifying assumptions. 
Here we present a finite volume extension of the SMM.

To have a more realistic model for finite volumes, we would like to account for the finite size  and geometrical shape of 
the largest fragments,  when  they  are comparable with the system volume.  
For this we will abandon the arbitrary size of largest fragment and
consider the constrained SMM (CSMM) in which
the largest fragment  size
is explicitly related to the volume $V$ of the system. 
A similar model, but with the fixed size of the largest fragment,  was 
recently analyzed in Ref. \cite{CSMM}.  

In this work  we will: 
solve the CSMM analytically at finite volumes using a new powerful method;   consider
how the first order phase transition 
develops from the singularities of the SMM isobaric partition \cite{Goren:81} in thermodynamic limit;
study   the  finite volume analogs of phases; and
discuss the finite size effects for  large fragments.


\vspace*{-0.25cm}

\section{Laplace-Fourier Transformation}

\vspace*{-0.25cm}

The system states in the SMM are specified by the multiplicity
sets  $\{n_k\}$
($n_k=0,1,2,...$) of $k$-nucleon fragments.
The partition function of a single fragment with $k$ nucleons is
\cite{Bondorf:95}:
$
V \phi_k (T) = V\left(m T k/2\pi\right)^{3/2}~z_k~
$,
where $k=1,2,...,A$ ($A$ is the total number of nucleons
in the system), $V$ and $T$ are, respectively, the  volume
and the temperature of the system,
$m$ is the nucleon mass.
The first two factors  on the right hand side (r.h.s.) 
of 
the single fragment partition 
originate from the non-relativistic thermal motion
and the last factor,
 $z_k$, represents the intrinsic partition function of the
$k$-nucleon fragment. Therefore, the function $\phi_k (T)$ is a phase space
density of the k-nucleon fragment. 
For \mbox{$k=1$} (nucleon) we take $z_1=4$
(4 internal spin-isospin states)
and for fragments with $k>1$ we use the expression motivated by the
liquid drop model (see details in \mbox{Ref. \cite{Bondorf:95}):}
$
z_k=\exp(-f_k/T),
$ with fragment free energy
\begin{equation}\label{one}
f_k = - W(T)~k 
+ \sigma (T)~ k^{2/3}+ (\tau + 3/2) T\ln k~,
\end{equation}
with $W(T) = W_{\rm o} + T^2/\epsilon_{\rm o}$.
Here $W_{\rm o}=16$~MeV is the bulk binding energy per nucleon.
$T^2/\epsilon_{\rm o}$ is the contribution of
the excited states taken in the Fermi-gas
approximation ($\epsilon_{\rm o}=16$~MeV). $\sigma (T)$ is the
temperature dependent surface tension parameterized
in the following relation:
$
\sigma (T)=\sigma_{\rm o}
[(T_c^2~-~T^2)/(T_c^2~+~T^2)]^{5/4},
$
with $\sigma_{\rm o}=18$~MeV and $T_c=18$~MeV ($\sigma=0$
at $T \ge T_c$). The last contribution in Eq.~(\ref{one}) involves the famous Fisher's term with
dimensionless parameter
$\tau$. 

The canonical partition function (CPF) of nuclear
fragments in the SMM
has the following form:
\begin{equation} \label{two}
\hspace*{-0.2cm}Z^{id}_A(V,T)=\sum_{\{n_k\}} \biggl[
\prod_{k=1}^{A}\frac{\left[V~\phi_k(T) \right]^{n_k}}{n_k!} \biggr] 
{\textstyle \delta(A-\sum_k kn_k)}\,.
\end{equation}
In Eq. (\ref{two}) the nuclear fragments are treated as point-like objects.
However, these fragments have non-zero proper volumes and
they should not overlap
in the coordinate space.
In the excluded volume (Van der
Waals) approximation
this is achieved
by substituting
the total volume $V$
in Eq. (\ref{two}) by the free (available) volume
$V_f\equiv V-b\sum_k kn_k$, where
$b=1/\rho_{{\rm o}}$
($\rho_{{\rm o}}=0.16$~fm$^{-3}$ is the normal nuclear density).
Therefore, the corrected CPF becomes:
$
Z_A(V,T)=Z^{id}_A(V-bA,T)
$.
The SMM defined by Eq. (\ref{two})
was studied numerically in Refs. \cite{Gupta:98,Gupta:99}.
This is a simplified version of the SMM, e.g. the symmetry and
Coulomb contributions are neglected.
However, its investigation
appears to be of  principal importance
for studies of the liquid-gas phase transition.

The calculation of $Z_A(V,T)$  
is difficult due to  the constraint $\sum_k kn_k =A$.
This difficulty can be partly avoided by 
evaluating
the grand canonical partition (GCP) 
%
\begin{equation}\label{three}
{\cal Z}(V,T,\mu)~\equiv~\sum_{A=0}^{\infty}
\exp\left({\textstyle \frac{\mu A}{T} }\right)
Z_A(V,T)~\Theta (V-bA) ~,
\end{equation}
where $\mu$ denotes a chemical potential.
The calculation of ${\cal Z}$  is still rather
difficult. The summation over $\{n_k\}$ sets
in $Z_A$ cannot be performed analytically because of
additional $A$-dependence
in the free volume $V_f$ and the restriction
$V_f>0 $.
This problem was resolved   \cite{Bugaev:00,Bugaev:01} 
by the Laplace transformation method to 
the so-called
isobaric ensemble \cite{Goren:81}.

In this work  we would like to  consider a more strict constraint
$\sum\limits_k^{K(V)} k~n_k =A$ , where the size
of the largest fragment  $K(V) = \alpha V/b$ cannot exceed the total volume of the system 
(the parameter $\alpha \le 1$  is introduced for convenience).
 The case $K(V) = Const$ is also included in our treatment. 
A similar restriction should be also applied to the upper limit of the product in 
all partitions $Z_A^{id} (V,T)$, $Z_A(V,T)$
and ${\cal Z}(V,T,\mu)$ introduced above 
(how to deal with the real values of $K(V)$, see later). 
Then the  model with this constraint, the CSMM,  cannot be solved by the Laplace 
transform method, because the volume integrals cannot be evaluated due to a complicated 
functional $V$-dependence.  
However, the CSMM can be solved analytically with the help of  the following identity 
\begin{equation}\label{four}
G (V) = 
%
\int\limits_{-\infty}^{+\infty} d \xi~ \int\limits_{-\infty}^{+\infty}
  \frac{d \eta}{{2 \pi}} ~ 
{\textstyle e^{ i \eta (V - \xi) } } ~ G(\xi)\,, 
\end{equation}
which is based on the Fourier representation of the Dirac $\delta$-function. 
The representation (\ref{four}) allows us to decouple the additional volume
dependence and reduce it to the exponential one,
which can be dealt by the usual Laplace transformation
in the  following sequence of steps
\begin{eqnarray} \label{five}
&&\hat{\cal Z}(\lambda,T,\mu)~\equiv ~\int_0^{\infty}dV~{\textstyle e^{-\lambda V}}
~{\cal Z}(V,T,\mu) = \nonumber\\
&&\hspace*{-0.1cm}\int_0^{\infty}\hspace*{-0.2cm}dV^{\prime}
\int\limits_{-\infty}^{+\infty} d \xi~ \int\limits_{-\infty}^{+\infty}
\frac{d \eta}{{2 \pi}} ~ { \textstyle e^{ i \eta (V^\prime - \xi) - \lambda V^{\prime} } } \times 
\nonumber \\
&& \sum_{\{n_k\}}\hspace*{-0.1cm} \left[\prod_{k=1}^{K( \xi)}~\frac{1}{n_k!}~\left\{V^{\prime}~
{\textstyle \phi_k (T) \,  
e^{\frac{ (\mu  - (\lambda - i\eta) bT )k}{T} }}\right\}^{n_k} \right] \Theta(V^\prime) = \nonumber \\
&&\hspace*{-0.1cm}\int_0^{\infty}\hspace*{-0.2cm}dV^{\prime}
\int\limits_{-\infty}^{+\infty} d \xi~ \int\limits_{-\infty}^{+\infty}
  \frac{d \eta}{{2 \pi}} ~ { \textstyle e^{ i \eta (V^\prime - \xi) - \lambda V^{\prime} 
+ V^\prime {\cal F}(\xi, \lambda - i \eta) } }\,.
%
%
%
\end{eqnarray}
After changing the integration variable $V \rightarrow V^{\prime} = V - b \sum\limits_k^{K( \xi)} k~n_k $,
the constraint of $\Theta$-function has disappeared.
Then all $n_k$ were summed independently leading to the exponential function.
Now the integration over $V^{\prime}$ in Eq.~(\ref{five})
can be straightforwardly done resulting in
\begin{equation}\label{six}
\hspace*{-0.4cm}\hat{\cal Z}(\lambda,T,\mu) = \int\limits_{-\infty}^{+\infty} \hspace*{-0.1cm} d \xi
\int\limits_{-\infty}^{+\infty} \hspace*{-0.1cm}
\frac{d \eta}{{2 \pi}} ~ 
\frac{  \textstyle e^{ - i \eta \xi }  }{{\textstyle \lambda - i\eta ~-~{\cal F}(\xi,\lambda - i\eta)}}~,
\end{equation}

\vspace*{-0.3cm}

\noindent
where the function ${\cal F}(\xi,\tilde\lambda)$ is defined as follows 
\begin{eqnarray}\label{seven}
&&\hspace*{-0.4cm}{\cal F}(\xi,\tilde\lambda) = \sum\limits_{k=1}^{K(\xi) } \phi_k (T) 
~e^{\frac{(\mu  - \tilde\lambda bT)k}{T} }
= \nonumber \\
&&\hspace*{-0.4cm}\left(\frac{m T }{2 \pi} \right)^{\frac{3}{2} } \hspace*{-0.1cm} \biggl[ z_1
~{\textstyle e^{ \frac{\mu- \tilde\lambda bT}{T} } } + \hspace*{-0.1cm} \sum_{k=2}^{K(\xi) }
k^{-\tau} e^{ \frac{(\mu + W - \tilde\lambda bT)k - \sigma k^{2/3}}{T} }  \biggr]\,.\,
\end{eqnarray}

As usual, in order to find the GCP by  the inverse Laplace transformation,
it is necessary to study the structure of singularities of the isobaric partition (\ref{seven}). 

\vspace*{-0.1cm}

\section{Isobaric Partition Singularities}

\vspace*{-0.2cm}

The isobaric partition (\ref{seven}) of the CSMM is, of course, more complicated
than its SMM analog \cite{Bugaev:00,Bugaev:01}
because for finite volumes the structure of singularities in the CSMM 
is much richer than in the SMM, and they match in the limit $V \rightarrow \infty$ only.
To see this let us first make the inverse Laplace transform:
\begin{eqnarray}\label{eight}
&&\hspace*{-0.6cm}{\cal Z}(V,T,\mu)~ = 
\int\limits_{\chi - i\infty}^{\chi + i\infty}
\frac{ d \lambda}{2 \pi i} ~ \hat{\cal Z}(\lambda,  T, \mu)~ e^{\textstyle   \lambda \, V } =
\nonumber \\
&&\hspace*{-0.6cm}\int\limits_{-\infty}^{+\infty} \hspace*{-0.1cm} d \xi
\int\limits_{-\infty}^{+\infty} \hspace*{-0.1cm}  \frac{d \eta}{{2 \pi}}  
\hspace*{-0.1cm} \int\limits_{\chi - i\infty}^{\chi + i\infty}
\hspace*{-0.1cm} \frac{ d \lambda}{2 \pi i}~ 
\frac{\textstyle e^{ \lambda \, V - i \eta \xi } }{{\textstyle \lambda - i\eta ~-~{\cal F}(\xi,\lambda - i\eta)}}~= 
\nonumber \\
&&\hspace*{-0.6cm}\int\limits_{-\infty}^{+\infty} \hspace*{-0.1cm} d \xi
\int\limits_{-\infty}^{+\infty} \hspace*{-0.1cm}  \frac{d \eta}{{2 \pi}}
\,{\textstyle e^{  i \eta (V - \xi)  } } \hspace*{-0.1cm} \sum_{\{\lambda _n\}}
e^{\textstyle  \lambda _n\, V } 
{\textstyle 
\left[1 - \frac{\partial {\cal F}(\xi,\lambda _n)}{\partial \lambda _n} \right]^{-1} } \,,
\end{eqnarray}
where the contour  $\lambda$-integral is reduced to the sum over the residues of all singular points
$ \lambda = \lambda _n + i \eta$ with $n = 1, 2,..$, since this  contour in the complex $\lambda$-plane  obeys the
inequality $\chi > \max(Re \{  \lambda _n \})$.  
Now both remaining integrations in (\ref{eight}) can be done, and the GCP becomes 
\begin{equation}\label{nine}
{\cal Z}(V,T,\mu)~ = \sum_{\{\lambda _n\}}
e^{\textstyle  \lambda _n\, V }
{\textstyle \left[1 - \frac{\partial {\cal F}(V,\lambda _n)}{\partial \lambda _n} \right]^{-1} } \,,
\end{equation}
i.e. the double integral in (\ref{eight}) simply  reduces to the substitution   $\xi \rightarrow V$ in
the sum over singularities. 
This is a remarkable result which 
can be formulated as the following 
\underline{\it theorem:}
{\it if the Laplace-Fourier image of the excluded volume GCP exists, then
for any additional $V$-dependence of ${\cal F}(V,\lambda _n)$ or $\phi_k(T)$
the GCP can be identically represented by Eq. (\ref{nine}).}


The simple poles in (\ref{eight}) are defined by the  equation 
\begin{equation}\label{ten}
\lambda _n~ = ~{\cal F}(V,\lambda _n)\,.
\end{equation}
In contrast to the usual SMM \cite{Bugaev:00,Bugaev:01} the singularities  $ \lambda _n $ 
are (i) 
 are volume dependent functions, if $K(V)$ is not constant,
and (ii) they can have a non-zero imaginary part, but 
in this case there  exist  pairs of complex conjugate roots of (\ref{ten}) because the GCP is real.

Introducing the real $R_n$ and imaginary $I_n$ parts of  $\lambda _n = R_n + i I_n$,
we can rewrite  Eq. (\ref{ten})
as a system of coupled transcendental equations 
\begin{eqnarray}\label{eleven}
&&\hspace*{-0.2cm} R_n = ~ \sum\limits_{k=1}^{K(V) } \tilde\phi_k (T)
~{\textstyle e^{\frac{Re( \nu_n)\,k}{T} } } \cos(I_n b k)\,,
\\
\label{twelve}
&&\hspace*{-0.2cm} I_n = - \sum\limits_{k=1}^{K(V) } \tilde\phi_k (T)
%
%
~{\textstyle e^{\frac{Re( \nu_n)\,k}{T} } } \sin(I_n b k)\,,
\end{eqnarray}
where we have introduced the 
set of  the effective chemical potentials  $\nu_n  \equiv  \nu(\lambda_n ) $ with $ \nu(\lambda) = \mu + W (T)  - \lambda b\,T$, and 
the reduced distributions $\tilde\phi_1 (T) = \left(\frac{m T }{2 \pi} \right)^{\frac{3}{2} }  z_1 \exp(-W(T)/T)$ and 
$\tilde\phi_{k > 1} (T) = \left(\frac{m T }{2 \pi} \right)^{\frac{3}{2} }  k^{-\tau}\, \exp(-\sigma (T)~ k^{2/3}/T)$ for convenience.


Consider the real root $(R_0 > 0, I_0 = 0)$, first. 
For $I_n = I_0 = 0$ the real root $R_0$ exists for any $T$ and $\mu$.
Comparing $R_0$ with the expression for vapor pressure of the analytical SMM solution 
\cite{Bugaev:00,Bugaev:01}
shows   that $T R_0$ is  a constrained grand canonical pressure of the gas. 
 As usual,  for  finite volumes the total mechanical pressure \cite{Hill}, as we will see in   section V, 
differs from   $T R_0$.
Equation (\ref{twelve}) shows that for $I_{n>0} \neq 0$ the inequality $\cos(I_n b k) \le 1$ never 
become the equality for all $k$-values  simultaneously. Then from Eq. (\ref{eleven})  
one obtains ($n>0$)
\begin{equation}\label{thirteen}
R_n < \sum\limits_{k=1}^{K(V) } \tilde\phi_k (T)
~{\textstyle e^{\frac{Re(\nu_n)\, k}{T} } } \quad \Rightarrow \quad R_n < R_0\,, 
\end{equation}
where the second inequality (\ref{thirteen}) immediately follows from the first one.
 In other words, the gas singularity is always the rightmost one.
This fact 
plays a decisive role in the thermodynamic limit
$V \rightarrow \infty$.

The interpretation of the complex roots $\lambda _{n>0}$  is less straightforward.
According to Eq. (\ref{nine}),   the  GCP is a superposition of  the
states of different  free energies $- \lambda _n V T$.  
(Strictly speaking,  $- \lambda _n V T$  has  a meaning of  the change of free energy, but
we  will use the  traditional  term for it.)
For $n>0$ the free energies  are complex. 
Therefore,  
 $-\lambda _{n>0} T$ is   the density of free energy.  The real part  of the free energy density,
  $- R_n T$, defines the significance
 of the state's  contribution to the partition:  due to (\ref{thirteen}) 
 the  largest contribution  always comes from the gaseous state and
 has the smallest  real part  of free energy density. As usual,  the states which do not have
 the smallest value of the (real part of)  free energy, i. e.  $- R_{n>0} T$, are thermodynamically metastable. 
 For  infinite   volume 
 they should not contribute unless they are infinitesimally close to  $- R_{0} T$, 
 but for finite volumes their contribution to the GCP may be important.

As one sees from (\ref{eleven}) and (\ref{twelve}), the states of  different  free energies have  
different values of the effective chemical potential $\nu_n$, which is not the case for
infinite volume \cite{Bugaev:00,Bugaev:01},
where there  exists a single value for the effective chemical potential. 
Thus,  for finite $V$
the  states which contribute to the GCP (\ref{nine}) are not in a true chemical equilibrium.

The meaning of the imaginary part of the free energy density  becomes  clear from 
(\ref{eleven}) and (\ref{twelve}): as one can see from (\ref{eleven})   the imaginary part $I_{n>0}$
effectively changes the number of degrees of freedom of  each $k$-nucleon fragment ($k \le K(V)$)
contribution to  the free energy  density  $- R_{n>0} T$.  It is clear, that the change 
of the effective number of degrees of freedom can occur virtually only and, if 
$\lambda _{n>0}$ state is accompanied  by 
some kind of  equilibration process. 
Both of these statements become clear,
 if we recall that  
the statistical operator in statistical mechanics and the  quantum mechanical convolution operator 
are related by the Wick rotation \cite{Feynmann}. In other words, the inverse temperature can be
considered as an  imaginary time.  
Therefore, depending on the sign,  the quantity  $ I_n b T \equiv \tau_n^{-1}$  that  appears 
 in the trigonometric  functions      
of  the  equations (\ref{eleven}) and (\ref{twelve}) in front of the imaginary time $1/T$ 
can be regarded  as the inverse decay/formation time $\tau_n$ of the metastable state which corresponds to the  pole $\lambda _{n>0}$ (for more details see next sections).
As will be shown further, for $\mu \rightarrow \infty$ the inverse chemical potential can be 
considered as a characteristic equilibration time as well.
 
This interpretation of  $\tau_n$ naturally explains the thermodynamic  metastability 
of all states except the gaseous one:
the metastable states can exist  in the system only virtually 
because of their finite decay/formation  time,
whereas the gaseous state is stable because it has an infinite decay/formation time.

%
%
\begin{figure}[ht]
\includegraphics[width=8.6cm,height=6.0cm]{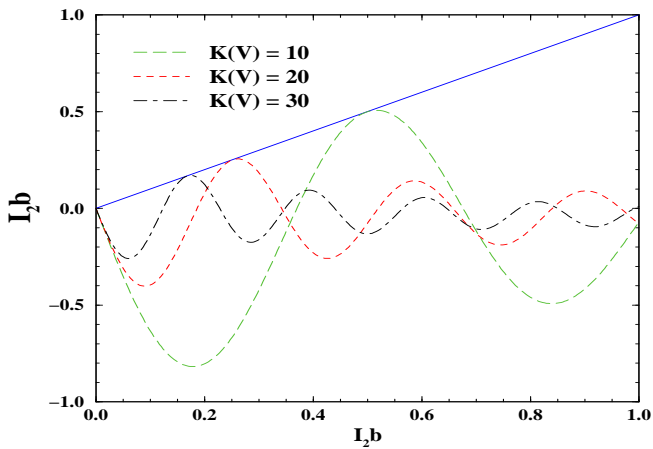}
\caption{A graphical solution of Eq. (\ref{twelve}) for $T = 10$ MeV and $\tau = 1.825$.
The l.h.s. (straight line) and  r.h.s. of Eq. (\ref{twelve}) (all dashed curves) are shown
as the function of
dimensionless parameter $I_1\,b$ for the three values of the largest fragment size $K(V)$.
The intersection point at $(0;\,0)$ corresponds to a real root of Eq. (\ref{ten}).
Each tangent point with the straight line generates  two complex  roots of (\ref{ten}).
}
  \label{fig1}
\end{figure}



\vspace*{-0.1cm}

\section{No Phase Transition Case}


\vspace*{-0.2cm}

It is instructive to treat the effective chemical potential $\nu (\lambda)$ as an independent variable
instead of $\mu$. In contrast to the infinite $V$, where  the upper   limit  $\nu \le 0$ defines the liquid phase singularity of the
isobaric partition and  gives the pressure of a liquid phase
$p_l(T,\mu) = T R_0 |_{V \rightarrow \infty}  = (\mu + W(T))/b$ \cite{Bugaev:00,Bugaev:01}, 
for finite  volumes and finite $K(V)$ the effective  chemical potential can
be complex (with either sign for its real part)  and its value defines the number and position of the imaginary roots 
$\{\lambda _{n>0} \}$ 
in the complex plane.
Positive  and negative values of the effective chemical potential  for finite systems  were  considered 
\cite{Elliott:01}
within the Fisher droplet model, but, to our knowledge,  its complex values have never  been  discussed.
From the definition of the effective chemical potential $\nu(\lambda)$ it is evident that  its complex
values for finite systems  exist  only  because of the excluded volume interaction, which is 
not taken into account in the  Fisher droplet model \cite{Fisher:67}.

As it is seen from  Fig.~1, the r.h.s. of Eq. (\ref{twelve})  
is the amplitude and frequency modulated sine-like  
function of dimensionless parameter $I_n\,b$. 
Therefore, depending on $T$ and $Re(\nu)$ values, there may exist 
no 
complex roots $\{\lambda _{n>0}\}$, a finite number of them, or an infinite number of them. 
In Fig.~1 we showed a special case which corresponds to  exactly three 
roots of Eq. (\ref{ten}) for each value of $K(V)$: the real root ($I_0 = 0$) and two complex conjugate
roots ($\pm I_1$). 
Since 
the r.h.s. of (\ref{twelve}) is monotonously increasing function
of  $Re(\nu)$, when the former is positive,  
it is possible to map the $T-Re(\nu)$ plane into
regions of a fixed number of roots of Eq. (\ref{ten}). 
Each curve in \mbox{Fig. 2} divides the $T-Re(\nu)$ plane
into three parts: for $Re(\nu)$-values below the curve there  is only one real root (gaseous phase), 
for points on  the curve there exist      
three roots, and above the curve there are five or more roots of Eq. (\ref{ten}).

For constant values of  $K(V) \equiv K$  the number of terms in the r.h.s. of (\ref{twelve}) does not depend on
the volume and, consequently, in thermodynamic limit $V \rightarrow \infty$   only the farthest
right simple pole in the complex $\lambda$-plane survives out of a finite number of simple poles.
According to the  inequality (\ref{thirteen}), the real root $\lambda_0$ is  the farthest  right singularity of isobaric partition (\ref{six}).
However,  there is a  possibility  that the real parts  of other  roots $\lambda_{n>0} $ become infinitesimally 
close to $R_0$, when there is an infinite number of terms which contribute to the GCP (\ref{nine}).

%
%
 \begin{figure}[ht]
  \includegraphics[width=8.6cm,height=6.0cm]{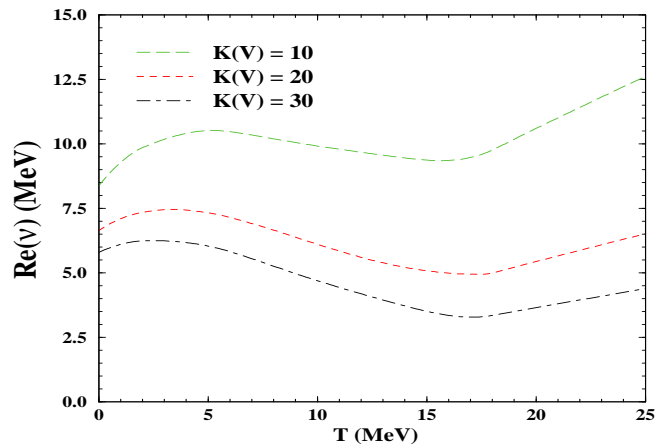}
  \caption{Each curve separates  the $T-Re(\nu_n)$ region of one real root of Eq. (\ref{ten})
(below the curve), three complex roots (at the curve) and five and more roots (above the curve)
for three values of $K(V)$ and the same parameters as in Fig. 1.
}
  \label{fig2}
\end{figure}

Let us show  now that even for an infinite number of simple poles in (\ref{nine})
only the real root $\lambda_0$  survives in the limit $V \rightarrow \infty$.
For this purpose consider  the limit
$Re(\nu_n)  \gg T $.
In this limit   the distance between  the imaginary parts of the nearest roots 
remains finite even for infinite volume.  Indeed,  for $Re(\nu_n)  \gg T   $
the leading contribution to the r.h.s. of (\ref{twelve}) corresponds to the harmonic with $k = K$,
and, consequently,  an exponentially large amplitude of this term
can be only  compensated by  a vanishing value of  $\sin\left( I_n \, b K  \right)$,  
i.e.   $I_n \, b K  =  \pi n + \delta_n$
with $|\delta_n| \ll \pi$ (hereafter we will analyze only 
the branch $I_n > 0$), 
and, therefore, the corresponding decay/formation time $\tau_n  \approx K [ \pi n T ]^{-1}$ 
is volume independent.

Keeping the leading term on the r.h.s. of  (\ref{twelve}) and solving for $\delta_n$, one finds
\begin{eqnarray}\label{Mfourteen}
\hspace*{-0.2cm}
I_n & \approx & (-1)^{n+1}  \tilde\phi_K (T)  ~{\textstyle e^{\frac{Re(\nu_n )\,K}{T} } } ~\delta_n \,, \\
\label{Mfifteen}
%
\delta_n &\approx  &  \frac{ (-1)^{n+1}  \pi n }{ K b ~ \tilde\phi_K (T)  }~{
\textstyle e^{- \frac{Re(\nu_n )\,K}{T} } } \,,    \\
\label{Msixteen}
R_n & \approx & (-1)^{n}  \tilde\phi_K (T)  ~{\textstyle e^{\frac{Re(\nu_n )\,K}{T} } }  \,, 
\end{eqnarray}
where in the last step we used Eq. (\ref{eleven}) and condition $|\delta_n| \ll \pi$.
Since for $V \rightarrow \infty$ all  negative values of $R_n$ cannot contribute to the 
GCP (\ref{nine}), it is sufficient to analyze even values of $n$ which, according to 
(\ref{Msixteen}),  generate  $R_n > 0$.

Since the inequality  (\ref{thirteen}) can not be broken,   a single possibility,
when $\lambda_{n>0}$ pole can contribute to the partition (\ref{nine}),  corresponds to 
the case 
$ R_n \rightarrow R_0 - 0^+$ 
for  some finite $n$.   
Assuming this, we find  $Re (\nu (\lambda_n)) \rightarrow Re(\nu (\lambda_0))$
for the same value of $\mu$. 

Substituting these results into equation (\ref{eleven}), one gets
\begin{equation}\label{Mseventeen}
\hspace*{-0.2cm}R_n \approx  \sum\limits_{k=1}^{K } \tilde\phi_k (T)
~{\textstyle e^{\frac{Re(\nu (\lambda_0) )\,k}{T} } } \cos\left[ \frac{ \pi n k}{K} \right]  \ll R_0\,.
\end{equation}
The inequality (\ref{Mseventeen})  follows  from the equation for $R_0$ and the fact that, even for
equal leading terms   in the sums above (with $k =  K$ and even  $n$),  the difference between $R_0$  and $R_n$ is  large due to  the next to leading term $k = K - 1$, which is  proportional to 
$e^{\frac{Re(\nu (\lambda_0) )\,(K-1)}{T} } \gg 1$.  
Thus, we arrive at  a  contradiction with our assumption $R_0 - R_n \rightarrow 0^+$, 
and, consequently,  it cannot be true.  Therefore,
for large volumes  the  real root $\lambda_0$  always gives
the main contribution to  the GCP (\ref{nine}), and  this is the only root that survives 
in the limit $V \rightarrow \infty$.
Thus,  we showed that  the model with the fixed  size  of the largest fragment has no phase transition because there is a single singularity of the isobaric partition (\ref{six}), which 
exists in thermodynamic limit.
 
\section{Finite Volume Analogs of Phases}
 
If $K(V)$ monotonically grows with the volume,  the situation is different. 
In this case for  positive value of $Re(\nu)  \gg T$ 
the leading exponent in the r.h.s. of (\ref{twelve})  
also corresponds to a largest fragment, i.e. to $k  = K(V)$. 
Therefore,  we can apply 
the same arguments  which were used above  for  the case $K(V) = K =  const$
and derive similarly  equations  (\ref{Mfourteen})--(\ref{Msixteen}) for  $I_n$ and $R_n$. 
From $I_n  \approx \frac{\pi n}{ b\, K( V) }$ it follows that,
when $V$ increases, the number of simple poles in (\ref{eight}) also increases 
and
the  imaginary part of 
the closest to the real $\lambda$-axis  poles becomes very small,
 i.e $I_n   \rightarrow 0$ for  $n \ll K(V)$,
 and, consequently, the associated   decay/formation time 
$\tau_n  \approx K(V)  [ \pi n T ]^{-1}$ grows with the volume of the system.
Due to $ I_n  \rightarrow 0$, 
the inequality (\ref{Mseventeen}) cannot be  
established for the poles with $n \ll K(V)$. 
Therefore, in  contrast to the previous case, for large $K(V)$ the  simple poles
with $n \ll K(V)$ will be infinitesimally close to the real axis of the complex $\lambda$-plane.


From Eq.  (\ref{Msixteen}) it follows that 
\begin{equation}\label{Meighteen}
R_n ~ \approx ~  \frac{p_l(T,\mu) }{T} -  \frac{ 1}{ K(V) b} 
 \ln \left|  \frac{ R_n}{  \tilde\phi_K (T)     }  \right|  \rightarrow  \frac{p_l(T,\mu) }{T} 
\end{equation}
for  $ | \mu | \gg T $ and $K(V) \rightarrow \infty$.   
Thus, 
we proved that
for infinite volume the  infinite number of simple poles moves toward 
the real $\lambda$-axis to the vicinity of liquid phase singularity $\lambda_l = p_l(T,\mu)/T $ 
of the isobaric partition
\cite{Bugaev:00, Bugaev:01} and
generates  an essential singularity of function ${\cal F}(V, p_l/T)$ in (\ref{seven}) 
{\it irrespective to the  sign of 
chemical potential $\mu$.}
As  we showed above, the states with $Re( \nu ) \gg T$  become  stable because they acquire   infinitely large 
decay/formation time $\tau_n$ in the limit $V \rightarrow \infty$.  
Therefore, these states should be identified 
as a liquid phase for finite  volumes as well. 
Such a conclusion can be easily understood, if we recall that
the  partial pressure $T R_n$  of  (\ref{Meighteen})
corresponds to a single  fragment of the largest possible size.

Now it is clear 
that each curve in Fig.~2  is   the  finite volume analog of the phase boundary $T-\mu$ for a given value of $K(V)$:  below the phase boundary there exists a gaseous phase, but at and above each curve there
are  states which can be identified with a finite volume analog of the mixed phase, and,
finally, at $ Re(\nu) \gg T$ there exists a liquid phase.
When  there is no phase transition, i.e. $K(V) = K = const$,  the structure of simple poles is
similar, but, first,  the line which separates the gaseous states from the metastable states does not
change with the volume, and, second, as shown above, the metastable states will never become
stable. 
Therefore,
a systematic study of the 
volume dependence  of free energy (or pressure for very large  $V$)  along with the formation and  decay times may  be  of a crucial importance for  experimental studies of 
the nuclear liquid gas phase transition.

The above results demonstrate that, in contrast  to Hill's expectations \cite{Hill}, the finite volume analog
of the mixed phase does not  
consist  just of  two pure phases. 
The mixed phase for finite volumes
consists of  a stable  gaseous phase and  the set of  metastable states which differ by the free energy. 
Moreover, the difference between the free energies of these states is  not surface-like, as Hill 
assumed in his treatment \cite{Hill}, but  volume-like.  Furthermore, 
according to Eqs. (\ref{eleven}) and (\ref{twelve}),  each of these states 
consists of the same fragments, but with different weights. 
As  seen above for the case $ Re(\nu) \gg T$,  
some fragments 
that  belong to 
the states, in which  the largest fragment is  dominant,
may, in principle, have negative weights (effective number of degrees of freedom) in 
the expression  for  $R_{n>0}$  (\ref{eleven}).
This can be understood easily because  higher concentrations of large fragments can be achieved 
 at the  expense of the  smaller fragments and  is reflected in  the corresponding change 
of the real part of  the free energy $- R_{n>0} V T$. 
Therefore, the  actual  structure of the mixed phase at finite volumes is  more complicated
than  was expected in earlier works.


A similar situation occurs  for the real values of $K(V)$. In this case all sums in 
Eqs. (\ref{ten}-\ref{thirteen}) should be expressed via the Euler-MacLaurin   
formula
\begin{equation*}
\hspace*{-0.2cm}\sum\limits_{k=1}^{K(V) } f_k = f(1) + \hspace*{-0.2cm}\int\limits_2^{K(V)}
\hspace*{-0.25cm} dk\, f(k) + 
{\textstyle \frac{f(K(V)) + f(2)}{2}  + \Delta_f (K) - }\,
\end{equation*}
%
\vspace*{-0.75cm}
\begin{equation} \label{fourteen}
\hspace*{-0.2cm}\Delta_f (2)\,, {\rm where} \quad \Delta_f (z) = \sum\limits_{n=1} \frac{B_{2n} }{(2n)!} 
~\frac{d^{2n} f(x)}{d\,x^{2n} }\biggr|_{x = z}\,.
%
%
\end{equation}
Here $B_{2n}$ are the Bernoulli numbers. 
The representation (\ref{fourteen}) 
allows one 
to study the effect of finite volume (FV) on the GCP (\ref{nine}). 

The above results are valid for any $K(V)$ dependence. However,  the linear one, i.e. 
$K(V) = \alpha V/ b$ with $0 < \alpha \le 1$, is the most natural.
With the help of the parameter $\alpha \le 1$ it is possible to describe a  difference 
between  the geometrical shape of the volume under consideration  and that one of the largest fragment.
For instance, by fixing $\alpha = \pi / 6$ it is possible to account for the fact that the largest
spherical  fragment  cannot fill completely  the cube with the side equal to its two radii, while 
there is enough space available for small fragments. 
Due to the $K(V) $ dependence 
in the CSMM there are  two different ways of how the finite volume affects thermodynamical functions: 
for finite $V$ and $Re(\nu_n)$ there is always a finite number of simple poles in (\ref{nine}), but their number and positions in the complex  $\lambda$-plane  depend on $V$.
To see this, let us study the mechanical pressure which corresponds to the GCP (\ref{nine})

\vspace*{-0.5cm}

\begin{eqnarray} \label{fifteen}
&&\hspace*{-0.7cm}p
= T \frac{\partial \ln {\cal Z}(V,T,\mu) }{\partial V} = 
{\textstyle \frac{T}{{\cal Z}(V,T,\mu) } } \sum\limits_{\{\lambda _n\}} \Biggl[ 
\frac{\lambda _n~ e^{\lambda _n \,V} }{ 1 - \frac{\partial {\cal F}(V,\lambda _n)}{\partial \lambda _n}  } 
+   \nonumber\\
&&\hspace*{-0.7cm}\frac{  e^{\lambda _n \,V} }{\left[ 
1 - \frac{\partial {\cal F}(V,\lambda _n)}{\partial \lambda _n} \right]^2  }
\Biggl\{ b^2\,\frac{ \partial \lambda _n}{\partial V}  \sum\limits_{k=1}^{K(V) } \tilde\phi_k \,k^2
~{\textstyle e^{\frac{\nu_n\,k}{T} } }  + \tilde\phi_{K(V)}  \times
\nonumber \\
&&\hspace*{-0.7cm}
{\textstyle e^{\frac{\nu_n\,K(V)}{T} } } 
{\textstyle K(V) 
\left[ 1 - \alpha +  \frac{\nu_n}{T} \left( \frac{1}{2} - \alpha \right) \right]}+  
o(K(V))\Biggr\} \Biggr] \hspace*{-0.05cm},
%
\end{eqnarray}

\vspace*{-0.3cm}

\noindent
where we give the main term for each $\lambda _n$ and  leading FV corrections explicitly for $Re(\nu_n)/T < 1$, whereas  
$o(K(V))$ accumulates the higher order corrections due to  
the  Euler-MacLaurin Eq. (\ref{fourteen}).
In evaluation of (\ref{fifteen}) we used an explicit representation of the
derivative $\partial \lambda _n/ \partial V$ which can be found from Eqs. (\ref{ten}) 
and (\ref{fourteen}). 
The first term in the r.h.s. of (\ref{fifteen}) describes the constrained grand canonical (CGC)
complex  pressure generated
by the simple pole $\lambda _n $ (due to its free energy density $- \lambda_n T$) weighted with the ``probability'' $e^{\lambda _n \,V }/{\cal Z} (V,T,\mu)$, 
whereas the second and third terms appear due to the volume dependence of $K(V)$. 
Note that, instead of the FV corrections, the usage of  natural values for 
$K(V)$ would generate the artificial delta-function   
terms in (\ref{fifteen}) for the volume derivatives.
Now it is clear  that in case  $K(V) = K = const$ the corrections to the  main term will  not appear,
and the number of poles and their positions will be defined by values of  $T$ and $\mu$ only.

As one can see from (\ref{fifteen}),
for finite volumes the corrections can give a non-negligible contribution to the pressure because
in this case 
$Re(\nu_n) > 0$ can be positive.   
The real parts of the partial CGC pressures $T \lambda _n$  may have either sign. 
Therefore, if the FV corrections to the pressure (\ref{fifteen}) are small, 
then,
according to  (\ref{thirteen}),  
the positive CGC pressures $T R_{n>0} > 0$ are mechanically  metastable 
and the negative ones $T R_{n>0} < 0$ 
are mechanically unstable compared to the gas pressure $T R_0$.
The FV corrections should be accounted for  to find the mechanically  meta- and unstable states 
in the general case. 
However, it is clear that the contribution of the  states with $T R_{n>0} < 0$  into partition and its derivatives
is exponentially small even for finite volumes.

As we showed earlier in this section,
when $V$ increases, the number of simple poles in (\ref{eight}) also increases and imaginary part of 
the closest to the real $\lambda$-axis  poles becomes very small. 
Therefore, for infinite volume the  infinite number of simple poles moves toward 
the real $\lambda$-axis to the vicinity of liquid phase singularity $p_l(T,\mu)/T$ and, thus,
generates  an essential singularity of function ${\cal F}(V, p_l/T)$ in (\ref{seven}). 
In this case the contribution of any of remote poles  from the real $\lambda$-axis  
to the GCP vanishes.  
Then it can be  shown that the FV corrections in (\ref{fifteen})
become negligible  because of the inequality $Re(\nu_n) \le 0$, and, consequently,
the reduced distribution of largest fragment 
$\tilde\phi_{K(V)} (T) = K(V)^{-\tau}\, \exp(-\sigma (T)~ K(V)^{2/3}/T)$
and the
derivatives $\partial \lambda _n/ \partial V$
vanish  for all $T$-values, and we obtain the usual SMM solution \cite{Bugaev:00,Bugaev:01}.
Its thermodynamics, as we discussed, 
is governed by the farthest right singularity in the complex $\lambda$-plane.

\vspace*{-0.1cm}

\section{Conclusions}

\vspace*{-0.2cm}

In this work we discussed  a  powerful mathematical method which allowed us to
solve analytically the CSMM at finite volumes. It is shown that  for finite volumes the GCP function 
can be identically rewritten in terms of the simple poles of the isobaric partition (\ref{six}).
The real pole $\lambda_0$  exists always and  the quantity  $T \lambda_0$ is the CGC pressure of the gaseous
phase.  The complex roots $\lambda_{n>0}$ appear as pairs of complex conjugate solutions
of equation (\ref{ten}).   As we discussed, their  most straightforward interpretation  is as follows: 
$- T Re (\lambda_{n>0}) $ has a meaning of 
the free energy density, whereas $ b T Im (\lambda_{n>0} )$, depending on  sign,  gives
the  inverse  decay/formation time  of such a state.
The gaseous state is always stable  because its decay/formation time is infinite and because
it has the smallest value of  free energy. 
The complex poles describe the metastable states for $Re( \lambda_{n>0} ) \ge  0 $ and mechanically
unstable states for $Re( \lambda_{n>0} ) < 0 $.

We studied the volume dependence of the simple poles and found a dramatic difference
in their  behavior  in case of phase transition and  without it.
For the former
this representation  allows one also to define the finite volume analogs of phases  unambiguously 
and to establish the finite volume analog of the $T-\mu$  phase diagram (see Fig. 2). 
At finite volumes the gaseous phase exists, if there is  a single simple pole, the mixed phase corresponds to three and more simple poles, 
whereas the liquid is represented by an infinite amount of simple poles at highest possible 
particle density (or $\mu \rightarrow \infty$).

As we showed, for given $T$ and $\mu$ the states of the mixed phase which have different  
$Re( \lambda_{n} )$ are
not in a true chemical equilibrium for finite volumes. This feature cannot be obtained within 
the Fisher droplet model due to lack
of the hard core repulsion between fragments.  This fact also  demonstrates clearly  that, 
in contrast to Hill's expectations \cite{Hill}, the mixed phase is  not just  a composition of two 
states which are the pure phases. 
As we showed, the mixed phase is a superposition of three and more collective states, 
and each of them 
is characterized by its own  value of $\lambda_n$.
Because of that  the difference between the  free energies  of these states is not a surface-like,
as Hill argued \cite{Hill}, but volume-like.

For the case with  phase transition, i.e.  for  $\frac{d K(V)}{d~ V} > 0$,  we analyzed what happens 
in thermodynamic limit. When $V$ grows,  the number of simple poles 
in (\ref{eight}) also increases and imaginary part of 
the closest to the real $\lambda$-axis  poles becomes vanishing. 
For infinite volume the  infinite number of simple poles moves toward 
the real $\lambda$-axis and forms  an essential singularity of function 
${\cal F}(V, p_l/T)$ in (\ref{seven}), which defines 
the  liquid phase singularity $p_l(T,\mu)/T$. Thus, we showed how
the phase transition develops in thermodynamic limit. 

Also 
we analyzed the finite volume corrections to the mechanical pressure (\ref{fifteen}). 
The corrections of a   similar kind should appear in the entropy, particle number and energy density
because of the $T$ and $\mu$ dependence of $\lambda _n$ due to (\ref{ten}) \cite{Bugaev:04a}. 
Therefore, these corrections should be taking into account while analyzing the experimental yields
of fragments. Then the phase diagram of the nuclear liquid-gas phase transition
can be recovered from the experiments on finite systems (nuclei)  with more confidence.  

A detailed analysis of the isobaric partition singularities in the $T - Re (\nu)$ plane allowed us
to define the finite volume analogs of  phases and  study  the behavior of these singularities  in the limit 
$V \rightarrow \infty$.  Such an analysis  opens a possibility  to study rigorously  the nuclear liquid-gas phase
transition  directly from the finite volume partition. This may help to extract 
the phase diagram of the nuclear liquid-gas phase transition 
from the experiments on finite systems (nuclei)  with more confidence.

{\bf  Acknowledgments.}
This work was supported by the US Department of Energy.
The fruitful  discussions with J. B. Elliott  and M. I. Gorenstein are appreciated.


\end{document}